\documentclass[runningheads]{llncs}
\usepackage[T1]{fontenc}
\usepackage{graphicx,amsmath,amssymb,mathrsfs,framed,esint,slashed,upgreek}
\usepackage[colorlinks]{hyperref}
\usepackage{color}
\newcommand{\rem}[1]{}

\newcommand{\de}{{\rm d}}

\newcommand{\bX}{{\mathbf{X}}}

\newcommand{\bfi}{\bfseries\itshape}

\def\contract{\makebox[1.2em][c]{\mbox{\rule{.6em}
{.01truein}\rule{.01truein}{.6em}}}}

\newcommand{\beq}{\begin{equation}}
\newcommand{\eeq}{\end{equation}}
\newcommand{\ben}{\begin{eqnarray}}
\newcommand{\een}{\end{eqnarray}}

\newcommand{\comment}[1]{\vspace{5mm}\par
\noindent\framebox{\begin{minipage}[c]{.98 \textwidth} \tt\bfi #1
\end{minipage}}\vspace{5 mm}\par}
\begin{document}
\title{Entropy functionals and equilibrium states\\in mixed quantum-classical dynamics}
\titlerunning{Mixed quantum-classical entropies}
%
\author{Cesare Tronci\inst{1} \and David Mart\'inez-Crespo\inst{2,3}
\and
Fran\c{c}ois Gay-Balmaz\inst{4}
}
\authorrunning{C. Tronci \it et al.}
%
 \institute{School of Mathematics and Physics, University of Surrey, Guildford, UK
\email{c.tronci@surrey.ac.uk}\\
 \and
Departamento de Matem\'aticas y Computaci\'on, Universidad de Burgos, Spain
\and
\makebox{Centro de Astropartículas y Física de Altas Energías, Universidad de Zaragoza, Spain}
\email{dmcrespo@ubu.es}\\
\and
Division of Mathematical Sciences, Nanyang Technological University, Singapore\\
\email{francois.gb@ntu.edu.sg}}
\maketitle              
\begin{abstract}
The computational challenges posed by many-particle quantum systems are often overcome by mixed quantum-classical (MQC) models in which certain degrees of freedom are treated as classical while others are retained as  quantum.  One of the fundamental questions raised by this hybrid picture involves the characterization of the information associated to MQC systems. Based on the theory of dynamical invariants in Hamiltonian systems, here we propose a family of hybrid entropy functionals that consistently specialize to the usual R\'enyi and Shannon entropies. Upon considering the MQC Ehrenfest model for the dynamics of quantum and classical probabilities, we apply the hybrid Shannon entropy  to characterize equilibrium configurations for simple Hamiltonians. The present construction also applies beyond Ehrenfest dynamics.

\keywords{Mixed quantum-classical dynamics  \and Hamiltonian structure  \and Casimir invariant \and Entropy functional \and Maximum-entropy principle.}
\end{abstract}
\section{Introduction: the mean-field model}

Mixed quantum-classical (MQC) models are especially well-known in computational chemistry and they go back to Born-Oppenheimer theory and its semiclassical approximation. Over the decades, hybrid quantum-classical formulations have also appeared in different fields, such as solid-state physics, spintronics, and, more recently, the theory of gravity. MQC models usually prescribe  the dynamics of a hybrid distribution-valued density matrix $\widehat{\cal P}(q,p)$ in such a way that ${D=\operatorname{Tr}\widehat{\cal P}}$ and ${\hat\varrho=\int \widehat{\cal P}\de q\de p}$ are the classical density and the quantum density-matrix, respectively. The dynamics of $\widehat{\cal P}(q,p)$ is prescribed in terms of the Hamiltonian matrix function $\widehat{H}(q,p)$, where $(q,p)$ are classical coordinates. Here, we exploit the dynamical invariants of Hamiltonian hybrid models to characterize MQC information and extend the  entropy  constructions from information theory.

In several cases, MQC models suffer from well-known consistency issues. The most common is the possibility for the hybrid density $\widehat{\cal P}$ to become unsigned over time, thereby  violating the Heisenberg principle and preventing the characterization of probability. On the one hand, these issues are absent in common mean-field models where $\widehat{\cal P}(q,p)=D(q,p)\hat\varrho$. In this case, the \makebox{equations read simply}
\[
\frac{\partial D}{\partial t}=\{{\operatorname{Tr}}(\hat\varrho\widehat{H}),D\}
\,,\qquad\qquad
i\hbar\frac{\de \hat\varrho}{\de t}=\left[{\int} D\widehat{H}\de q\de p,\hat\varrho\right].
\]
On the other hand, such models neglect correlation effects thereby leading to trivial dynamics of the \emph{quantum purity}  $\operatorname{Tr}\hat\varrho^2$. In realistic cases, the latter  undergoes nontrivial evolution which is commonly referred to as \emph{quantum decoherence} \cite{Subotnik}. Despite  important limitations, the clear identification of  quantum and classical probabilities in  the mean-field context allows \makebox{writing the hybrid MQC entropy as}
\beq\label{MFEnt}
S(\widehat{\cal P})=-{\operatorname{Tr}}{\int} \widehat{\cal P}\ln \widehat{\cal P}\,\de q\de p=-{\operatorname{Tr}}(\hat\varrho\ln \hat\varrho)-{\int} D\ln D\,\de q\de p,
\eeq
that is the sum of the quantum von Neumann entropy and the classical Shannon entropy. As a functional of the type $\int\Upgamma(D,\hat\varrho)\de q \de p$, for any real-valued analytic function $\Upgamma$, the quantity \eqref{MFEnt} is conserved by the reversible mean-field MQC dynamics and may be used to characterize the quantum-classical information following standard procedures \cite{Alonso}. A R\'enyi entropy is also available in the form ${\cal H}_\alpha=(\ln\operatorname{Tr}\hat\varrho^\alpha+\ln \int \!D^\alpha\,\de q\de p)/(1-\alpha)$, so that \eqref{MFEnt} is recovered in the limit $\alpha\to1$. This simple situation, however, is accompanied by the long-standing \emph{detailed balance problem} \cite{Tully}. In particular, no explicit equilibrium profile is made available by the standard Maximum Entropy principle 
\begin{multline}
\delta\left[ {\operatorname{Tr}}{\int} D\hat\varrho\ln (D\hat\varrho)\de q\de p+\mu{\left({\operatorname{Tr}}{\int} D\hat\varrho\widehat{H}\de q\de p-E\right)}\right.
\\
\left.
+\lambda_1({\operatorname{Tr}}\hat\varrho-1)+{\lambda_2}{\left({\int} D\de q\de p-1\right)}\right]=0.
\end{multline}
Rather, this yields ${1+\lambda_1+\ln\hat\varrho+\mu{\int} D\widehat{H}\de q\de p=0}$ and $1+\lambda_2+\ln D+{\operatorname{Tr}}(\hat\varrho\widehat{H})=0$, 
which are hardly solved  beyond the uncoupled case ${\widehat{H}(q,p)=H_C(q,p)\boldsymbol{1}}+\widehat{H}_Q$.

Despite this challenging point, the identification of quantum-classical entropies remains straightforward in the mean-field case. This situation changes drastically when one tries  to capture quantum-classical correlations by going beyond the mean-field factorization $\widehat{\cal P}(q,p)=D(q,p)\hat\varrho$. In this case, the quantity $-{\operatorname{Tr}}{\int} \widehat{\cal P}\ln \widehat{\cal P}\,\de q\de p$ in \eqref{MFEnt} generally fails to be preserved by the reversible dynamics so that the second law of thermodynamics is violated. In order to overcome this difficulty, we propose to identify suitable entropy functionals by resorting to the Hamiltonian structure (where available) of the underlying MQC model: since entropy must be conserved for arbitrary Hamiltonians, it has to be a Casimir for the Poisson bracket associated to the model under consideration. In order to avoid the important issues that may emerge in the case of an infinite-dimensional quantum Hilbert space $\mathscr{H}$, here we will consider the finite-dimensional case $\mathscr{H}=\Bbb{C}^n$. 

The remainder of this paper  focuses on the \emph{Ehrenfest model}, which underlies the multi-trajectory Ehrenfest scheme commonly adopted in MQC molecular dynamics \cite{TrGB23}. While this model is accompanied by relevant accuracy issues, here we use it as a basis for our construction. As discussed in Section \ref{beyond}, the latter also applies in more advanced contexts beyond Ehrenfest dynamics. 

\section{The Ehrenfest model and its entropy functionals}
The Ehrenfest model is  written in terms of the hybrid density \makebox{operator $\widehat{\cal P}(q,p)$ as}
\beq\label{EhrMod1}
i\hbar\partial_t \widehat{\cal P}+i\hbar\operatorname{div}(\widehat{\cal P}\langle\bX_{\widehat{H}}\rangle)=[\widehat{H},\widehat{\cal P}].
\eeq
Here, we have used the notation $\langle{\widehat{A}}\rangle=
\operatorname{Tr}(\widehat{\cal P}\widehat{A})/\operatorname{Tr}\widehat{\cal P}$, for any operator-valued function $\widehat{A}(q,p)$, while $\bX_{\widehat{H}}=(\partial_p\widehat{H},-\partial_q\widehat{H})$ is the hybrid Hamiltonian vector field.
Equation \eqref{EhrMod1} is Hamiltonian with the non-canonical Poisson bracket structure \cite{JGM}
\begin{equation}\label{bracket_candidate_rho}
\{\!\!\{f,h\}\!\!\}(\widehat{\cal P})=\int \!\bigg(\frac1{\operatorname{Tr}\widehat{\cal P}  }\bigg(\widehat{\cal P}  {:} \bigg\{\frac{\delta f}{\delta \widehat{\cal P}},\frac{\delta h}{\delta \widehat{\cal P}}\bigg\}{:} \widehat{\cal P}   \bigg)
 -\left\langle \widehat{\cal P}  ,\frac{i}\hbar\!\left[\frac{\delta f}{\delta \widehat{\cal P}},\frac{\delta h}{\delta \widehat{\cal P}}\right] \right\rangle\bigg)\de q\de p,
\end{equation}
and Hamiltonian functional 
\[
h(\widehat{\cal P})=\operatorname{Tr}\int\widehat{\cal P}\widehat{H}\,\de q\de p.
\]  
The notation is such that the operation $A{:}B=\operatorname{Tr}(AB)$ has priority. Also, $\langle A,B\rangle={\operatorname{Re}}{\operatorname{Tr}}(A^\dagger B)$ defines the real-valued pairing, while $\{\cdot,\cdot\}$ is the canonical Poisson bracket on phase-space. 

If instead the  Hamiltonian functional $h(\widehat{\cal P})$ is left arbitrary, the Hamiltonian equation associated to \eqref{bracket_candidate_rho} is found by  $\dot{f}=\{\!\!\{f,h\}\!\!\}$ and reads 
\beq
i\hbar\frac{\partial\widehat{\cal P}}{\partial t}+i\hbar\operatorname{div}\!\Big(\widehat{\cal P}\big\langle \bX_{{\delta h}/{\delta \widehat{\cal P}}}\big\rangle\Big)=\Big[\frac{\delta h}{\delta \widehat{\cal P}},\widehat{\cal P}\Big].
\label{HamEqn}
\eeq
As a result,  the Poisson bracket \eqref{bracket_candidate_rho} is seen to possess the Casimir invariant \cite{JGM}
\beq\label{Casimir1}
C_1(\widehat{\cal P})=\operatorname{Tr}\int \!\widehat{\cal P}\,\Phi\bigg(\frac{\widehat{\cal P}}{\operatorname{Tr}\widehat{\cal P}}\bigg)\,\de q\de p,
\eeq
for any analytic function $\Phi:\operatorname{Her}(n)\to\Bbb{R}$, where $\operatorname{Her}(n)$ denotes the space of $n$-dimensional Hermitian matrices. The functional $C_1$ is a Casimir  in the sense that $\{\!\!\{f,C_1\}\!\!\}=0$ for any functional $f(\widehat{\cal P})$. 

The choice $\Phi(\widehat{A})=-\operatorname{Tr}(\widehat{A}\ln\widehat{A})$ yields the functional $- {\operatorname{Tr}}{\int} \widehat{\cal P}\ln({\widehat{\cal P}}/{\operatorname{Tr}\widehat{\cal P}})\de q\de p$, which crucially differs  from the expression appeared after the first equality in \eqref{MFEnt}. Importantly, this expression fails to recover the  mean-field entropy \eqref{MFEnt} in the case $\widehat{\cal P}(q,p)=D(q,p)\hat\varrho$. Thus, extra invariants are needed in order to provide a complete characterization of the overall MQC entropy in Ehrenfest dynamics.

\subsection{Conditional pure-state representation}
More insight can be obtained by considering a hybrid density operator  of the form $\widehat{\cal P}(q,p)=\Upsilon(q,p)\Upsilon(q,p)^\dagger$, or, equivalently, 
\[
\widehat{\cal P}(q,p)=D(q,p)\psi(q,p)\psi(q,p)^\dagger.
\] 
Here, we wrote $\Upsilon=\sqrt{D}\psi$, where $\psi$ is a \emph{conditional state vector}, so  that $\|\psi(q,p)\|^2\linebreak=1$ and $\|{\cdot}\|$ is the norm on the quantum Hilbert space $\mathscr{H}=\Bbb{C}^n$. This representation of the hybrid density has the advantage of splitting the classical density from the conditional quantum dynamics. As we will see, a further advantage is that it leads to an additional family of dynamical invariants.

In this representation, the chain rule relation
\[
\frac{\delta h}{\delta \widehat{\cal P}}\psi=
\frac{\delta h}{\delta D}\psi-\frac1{2D}\Big\langle\frac{\delta h}{\delta \psi},\psi\Big\rangle\psi+\frac1{2D}\frac{\delta h}{\delta \psi}
\]
takes \eqref{HamEqn} into the system
\beq\label{D_psi_equ}
\partial_t D+\operatorname{div}(D\boldsymbol{\cal X})=0
,\ \qquad\ 
i\hbar(\partial_t+\boldsymbol{\cal X}\cdot\nabla)\psi=\frac1{2D}\frac{\delta h}{\delta \psi}
,
\eeq 
with
\[
\boldsymbol{\cal X}=\bX_{\textstyle\frac{\delta h}{\delta D}}-\frac1{D}\Big\langle\frac{\delta h}{\delta \psi},\bX_\psi\Big\rangle.
\]
Here, the real valued pairing $\langle\cdot{,}\cdot\rangle$ is given by the real part of the inner product $\langle\psi_1|\psi_2\rangle=\psi_1^\dagger\psi_2$, that is $\langle\cdot{,}\cdot\rangle=\operatorname{Re}\langle\cdot|\cdot\rangle$ and 
$h(\psi)={\int} D\langle\psi,\widehat{H}\psi\rangle\de q \de p$, so that $\boldsymbol{\cal X}=\langle \bX_{\widehat{H}}\rangle$.
We observe that, while the classical density is transported by the vector field $\boldsymbol{\cal X}$, the conditional state evolves unitarily while being swept in the phase-space frame moving with $\boldsymbol{\cal X}$. Importantly, the latter vector field is neither Hamiltonian nor incompressible in general; as a result, the usual Shannon entropy ${-}{\int} D\ln D\de q\de p$ fails to be an invariant of motion. Thus, as anticipated, the entropy functionals for Ehrenfest dynamics must involve extra features.

In the present representation, using the second equation in \eqref{D_psi_equ} shows that the Berry connection $\boldsymbol{\cal A}_B=\langle\psi,-i\hbar\nabla\psi\rangle$ satisfies the relation
\[
(\partial_t+\pounds_{\boldsymbol{\cal X}})\boldsymbol{\cal A}_B=-\boldsymbol{\cal X}\contract \omega+\nabla\Big(\frac{\delta h}{\delta D}-\frac1{2D}\Big\langle\frac{\delta h}{\delta \psi},\psi\Big\rangle\Big),
\]
where $\omega=\de q\wedge\de p$ is the canonical symplectic form and, in components, $(\boldsymbol{\cal X}\contract \omega)_k\linebreak=\boldsymbol{\cal X}^j \omega_{jk}$, so that 
$
\boldsymbol{\cal X}\contract \omega=\nabla({\delta h}/{\delta D})-D^{-1}\langle{\delta h}/{\delta \psi},\nabla\psi\rangle$.
Also, 
 $\pounds_{\boldsymbol{\cal X}}$ denotes the Lie derivative, in this case applied to the Berry connection one-form, so that the product rule gives $\pounds_{\boldsymbol{\cal X}}\boldsymbol{\cal A}_B=\langle({\boldsymbol{\cal X}}\cdot\nabla\psi),-i\hbar\nabla\psi\rangle+\langle\psi,-i\hbar\nabla({\boldsymbol{\cal X}}\cdot\nabla\psi)\rangle$. Then, we recognize that, by Cartan's formula, $-\boldsymbol{\cal X}\contract \omega=\pounds_{\boldsymbol{\cal X}}\boldsymbol{\cal A}-\nabla({\boldsymbol{\cal X}}\cdot\boldsymbol{\cal A})$, where $\boldsymbol{\cal A}=(p,0)$ are the phase-space components of the canonical one-form on phase-space ${\cal A}=pdq$. Since, the latter is constant in time, we obtain the relation
\beq\label{oneform}
(\partial_t+\pounds_{\boldsymbol{\cal X}})(\boldsymbol{\cal A}-\boldsymbol{\cal A}_B)=\nabla\Big({\boldsymbol{\cal X}}\cdot\boldsymbol{\cal A}+\frac1{2D}\Big\langle\frac{\delta h}{\delta \psi},\psi\Big\rangle-\frac{\delta h}{\delta  D}\Big),
\eeq
which unfolds the MQC Poincar\'e integral invariant, that is 
\[
{\oint_{c_t}}\Big(\big\langle\psi,(p+i\hbar\partial_q)\psi\big\rangle\,\de q+\big\langle\psi,i\hbar\partial_p\psi\big\rangle\,\de p\Big)=const.,
\]
where $c_t$ is a phase-space loop moving with the flow of $\boldsymbol{\cal X}$. More importantly, taking the differential of \eqref{oneform}, we have
\[
(\partial_t+\pounds_{\boldsymbol{\cal X}})(\omega+{\cal B})=0
\] 
where
${\cal B}=\de\boldsymbol{\cal A}_B=\hbar\operatorname{Im}\{\psi^\dagger,\psi\}\omega$
is the Berry curvature. Thus, upon introducing the \emph{Liouville volume} 
\[
\Lambda=1+\hbar\operatorname{Im}\{\psi^\dagger,\psi\},
\]
such that $\omega+{\cal B}=\Lambda\omega$, the latter is a symplectic form at all times if it is so initially (although the flow of $\boldsymbol{\cal X}$ is not generally symplectic). Also, in the general case of a $2N$-dimensional phase-space,  the wedge power $(\omega+{\cal B})^{\wedge N}$ identifies a  Lie-transported Liouville volume form $\Lambda$, that is
\[
\partial_t\Lambda+\operatorname{div}(\Lambda\boldsymbol{\cal X})=0.
\]
For simplicity,  we will restrict to   a two-dimensional phase-space.

At this point,  it becomes clear that any functional of the type \cite{GBTr22a}
\beq\label{Casimir2}
C_2(D,\psi)={\int} D\,\Sigma{\bigg(\frac{\Lambda}D\bigg)}\de q\de p
\eeq
identifies a dynamical invariant for any function $\Sigma:\Bbb{R}\to\Bbb{R}$. Note that, unlike $C_1$ in \eqref{Casimir1}, $C_2$  depends on the derivative of $\psi$, so the two invariants have very different nature. Now, if we let $\Sigma(x)=\ln x$, then we are led to the following entropy functional:
\beq\label{ent1}
S(D,\psi)=-\int\!D\ln\frac{D}{\Lambda}\,\de q\de p.
\eeq
We remark that this functional may be obtained from the MQC R\'enyi entropy ${\cal H}_\alpha=({1-\alpha})^{-1}\log \int\! \Lambda ({D}/\Lambda)^\alpha\,\de q\de p$ in the limit $\alpha\to1$. This construction unfolds the duality between the Liouville volume $\Lambda$ and the scalar function $D/\Lambda$, with the former playing the role of the integration measure and the latter behaving as a probability distribution function. We also note that \eqref{ent1} is (minus) the Kullback-Leibler divergence of $D$ from $\Lambda$, while the more general expression given by $C_2$ is  referred to as the $\Sigma$-divergence of $\Lambda$ from $D$, for $\Sigma: \mathbb{R}_{\geq0}\rightarrow\mathbb{R}$ a convex function with $\Sigma(1)=0$. The entropy \eqref{ent1} has a direct counterpart in the physics of guiding-center plasmas \cite{BurbyTronci}.

 We observe that the additional denominator, while carrying information on the quantum-classical correlations, is also necessary to reflect the fact that the classical vector field $\boldsymbol{\cal X}$ is not incompressible. In addition, we also notice that the volume $\Lambda=1+\hbar\operatorname{Im}\{\psi^\dagger,\psi\}$ is not generally sign-definite. However, the requirement of a positive-definite $\Lambda$ can be set as an initial condition as the Lie-transport of a volume form preserves its sign. For example, an easy situation is given by initial conditions such that ${\operatorname{Im}\{\psi^\dagger,\psi\}=0}$, which is satisfied if there exists a function $\zeta(q,p)$ such that ${\psi=\phi\circ\zeta}$, for some state vector $\phi$ depending on  one-parameter. Here, the symbol $\circ$ denotes standard composition of functions. Also, the Berry curvature  vanishes if $\psi$ is purely real, or purely imaginary. In all these cases,  \eqref{ent1} reduces to the classical Shannon entropy.

While the functional \eqref{ent1} and its R\'enyi generalization ${\cal H}_\alpha$ appear as the most obvious MQC entropy candidates, we see that \eqref{ent1} fails to recover  the entire  mean-field entropy  \eqref{MFEnt} unless the latter is specialized to $\hat\varrho=\uppsi\uppsi^\dagger$, where $\uppsi$ is a constant state vector. A more general representation overcoming this issue is found below.

\subsection{Uhlmann representation of conditional density matrices}

A convenient method to proceed beyond the conditional pure-state representation is to resort to the \emph{Uhlmann representation} \cite{Uhlmann}. Originally appeared as a convenient representation for mixed states, this consists in writing a general $n$-dimensional density matrix as $\hat\varrho=WW^\dagger$, where $W{\in \Bbb{C}^{n\times m}}$ is some rectangular matrix also known as \emph{wave operator} \cite{Bondar}. Notice that $W$ is only defined up to the right multiplication by an  arbitrary $m$-dimensional unitary matrix, which represents a non-Abelian gauge choice. The evolution resulting from the quantum Liouville equation $i\hbar\,\de\hat\varrho/\de t=[\widehat{H},\hat\varrho]$ is $i\hbar\dot{W}=\widehat{H}W$, that is $W(t)=e^{-i\widehat{H}/\hbar}W_0$. In the context of symplectic geometry, as explained in \cite{Skerritt}, the quantity $-i\hbar^{-1}WW^\dagger$ comprises a momentum map structure for the left multiplication of $n$-dimensional unitary matrices on the space $\Bbb{C}^{n\times m}$ of wave operators. This space is endowed with the symplectic form $\Omega(W_1,W_2)=2\hbar\operatorname{Im}\langle W_1|W_2\rangle$, where we use the Frobenius inner product $\langle W_1|W_2\rangle=\operatorname{Tr}(W_1^\dagger W_2)$. Alternatively, $W^\dagger W$ is the Noether conserved quantity for the gauge symmetry given by the right multiplication.

In the MQC setting, wave operators may be used to introduce the representation $\widehat{\cal P}(q,p)= {\cal W}(q,p){\cal W}(q,p)^\dagger$ or, equivalently, 
\[
\widehat{\cal P}(q,p)=D(q,p) W(q,p)W(q,p)^\dagger.
\] 
Here, we wrote ${\cal W}=\sqrt{D} W$, where  $W$ is a \emph{conditional wave operator}, so  that  $\|W(q,p)\|^2=1$ and $\|{\cdot}\|$ is the Frobenius norm. Similarly, $\hat\rho(q,p)=W(q,p)W(q,p)^\dagger$ is a \emph{conditional density matrix} not to be confused with the density matrix $\hat\varrho$ of the quantum subsystem. We remark that this Uhlmann representation comprises all possible hybrid operators.   In this case, the chain rule 
\[
\frac{\delta h}{\delta \widehat{\cal P}}W=
\frac{\delta h}{\delta D}W-\frac1{2D}\Big\langle\frac{\delta h}{\delta W},W\Big\rangle W+\frac1{2D}\frac{\delta h}{\delta W}
\]
takes \eqref{HamEqn} into the system
\beq\label{HamEqn2}
\partial_t D+\operatorname{div}(D\boldsymbol{\cal X})=0
,\ \qquad\ 
i\hbar(\partial_t+\boldsymbol{\cal X}\cdot\nabla )W=\frac1{2D}\frac{\delta h}{\delta 
W}
,
\eeq 
with
\[
\boldsymbol{\cal X}=\bX_{\textstyle\frac{\delta h}{\delta D}}-\frac1{D}\Big\langle\frac{\delta h}{\delta W},\bX_W\Big\rangle.
\]
Here, the Hamiltonian functional is given as $h(D,W)={\int}D{\langle W|\widehat{H}W\rangle}{\de} q{\de} p$ and  we have used the  pairing notation ${\langle\cdot{,}\cdot\rangle=\operatorname{Re}\langle \cdot|\cdot\rangle}$. At this point, all the previous  steps go through in exactly the same fashion as in the previous case upon redefining the Berry connection as $\boldsymbol{\cal A}_B=\langle W,-i\hbar\nabla W\rangle$. Then, the Hamiltonian equations \eqref{HamEqn2} possess the following family of dynamical invariants:
\beq\label{Casimir}
C(D,W)={\int} D\Gamma{\Big(WW^\dagger,\frac{\Lambda}D\Big)}\de q\de p
\eeq
for any analytic function $\Gamma:\Bbb{C}^{n\times n}\times\Bbb{R}\to\Bbb{R}$. This expression extends  both $C_1$ in \eqref{Casimir1} and $C_2$, which are recovered by $\Gamma(\widehat{A},x)=\Phi(\widehat{A})$ and  $\Gamma(\widehat{A},x)=\Sigma(x)$, respectively. Notice that, in this case, the Liouville volume form reads $\Lambda=\omega+{\cal B}$, where ${\cal B}=\de\boldsymbol{\cal A}_B=\hbar{\operatorname{Im}\operatorname{Tr}}{\{W^\dagger{,}W\}}\omega$. The invariants \eqref{Casimir} lead us to making the choice $\Gamma(\widehat{A},x)=-\langle\widehat{A},\ln\widehat{A}\rangle+\ln x$, thereby leaving us with the following entropy functional:
\beq\label{ent2}
S(D,W)=-{\operatorname{Tr}}{\int}{\widehat{\cal P}}\ln\frac{\widehat{\cal P}}{\Lambda}\, \de q\de p=-{\int} \Big\langle\frac{D}{\Lambda} WW^\dagger,{\ln}\Big(\frac{D}{\Lambda}WW^\dagger \Big)\Big\rangle\,\Lambda\de q\de p,
\eeq
where we emphasize that $\widehat{\cal P}$ is written in terms of $W$, whose gradients appear explicitly in $\Lambda$.
Once again, this functional may be obtained  as the limit $\alpha\to1$ of the MQC R\'enyi entropy 
\begin{equation}\label{MQCRenyi}
{\cal H}_\alpha=\frac1{1-\alpha}{\ln} {\int} \Lambda\, {\operatorname{Tr}}{\bigg(\frac{\widehat{\cal P}}{\Lambda}\bigg)^{\!\alpha}}\de q\de p
=\frac\alpha{1-\alpha}\ln  \bigg\|\frac{DWW^\dagger}\Lambda\bigg\|_\alpha^\Lambda
,
\end{equation}
where we define ${\|\widehat{A}\|_\alpha^\Lambda=(\int (\|\widehat{A}\|_\alpha)^\alpha\Lambda\de q\de p)^{1/\alpha}}$ and $\|\widehat{A}\|_\alpha=(\operatorname{Tr}\widehat{A}^\alpha)^{1/\alpha}$ is the Schatten  norm of a positive-semidefinite matrix $\widehat{A}$. This invariant functional arises as ${(1-\alpha)^{-1}}\ln C$, where $C$ is given as in \eqref{Casimir} by setting $\Gamma(\widehat{A},x)=x^{1-\alpha}{\operatorname{Tr}}{\widehat{A}^\alpha}$. We observe that  the MQC entropies above reduce respectively to the Shannon mean-field entropy \eqref{MFEnt} and its R\'enyi extension in the case $\nabla W=0$.

\section{Maximum entropy principle and equilibrium states}
Having characterized the analogues of the Shannon and R\'enyi entropies for MQC Ehrenfest dynamics, we apply this construction to characterize maximal-entropy equilibrium states by using Jaynes' maximum-entropy principle. 

We will first proceed in the conditional pure-state representation by considering the following variational problem:
\begin{multline*}
\delta\left[ {\int} D\ln (D/\Lambda)\de q\de p+\mu{\left({\int} D\langle\psi,\widehat{H}\psi\rangle\de q\de p-E\right)}\right.
\\
\left.
+{\int}D\lambda_1(\|\psi\|^2-1)\de q\de p+{\lambda_2}{\left({\int} D\de q\de p-1\right)}\right]=0.
\end{multline*}
By a slight abuse of the dot product notation in the relation $\delta \Lambda
  {=}
  -2
  \langle
    \nabla\delta \psi
    , \cdot
    i\hbar \bX_{\psi}
  \rangle$, taking variations leads to the following equations:
\[
      \Lambda{(
      \mu\widehat{H}
      +
        {\lambda_1}
      )}
      \psi=
-i\hbar
      \{{\ln}{(\Lambda^{-1}{D})} ,\psi 
      \}
,\qquad\quad
\mu\langle
  \psi
  ,
  \widehat{H}
  \psi
  \rangle
+
1+
{\ln}{(\Lambda^{-1}{D})}
+{\lambda_2}=0.
\]
Taking into account the normalization of $D$, the second condition yields
\[
\frac{D}\Lambda= \frac{e^{-  \mu\langle
    \widehat{H}
    \rangle}}{Z_C}
    \,,\qquad\quad 
    Z_C={\int}\Lambda e^{- \mu \langle
    \widehat{H}\rangle}\de q \de p,
\]
while the first one becomes 
\beq\label{marina}
\Lambda{(
      \widehat{H}
      +
        \mu^{-1}{\lambda_1}
      )}
      \psi=
      -i\hbar
      \{{\langle\psi,\widehat{H}\psi\rangle} ,\psi 
      \},
\eeq
which is a formidable equation that is solved here in two simple cases.

In the first case, the Hamiltonian depends on the phase-space coordinates only through a function $\zeta(q,p)$, so that $\widehat{H}=\widehat{H}\circ\zeta$. Then, we observe that \eqref{marina} is solved by the eigenvectors $\psi_n(\zeta)$ of $\widehat{H}(\zeta)$, so that $\Lambda=1$  and one is left with
\[
\big(\widehat{H}+
        {\cal E}_n
      \big)
      \psi_n=0
      \,,\qquad\qquad
      D_n= \frac{e^{-  \mu{\cal E}_n}}{{\int} e^{- \mu {\cal E}_n}\de q \de p},
\]
so that  the available classical equilibria are labelled by an integer value corresponding to the eigenvalue ${\cal E}_n$. For example, equilibria of this type are available for  MQC Hamiltonians of the type $\widehat{H}=m^{-1}p^2/2+\eta p\widehat{\sigma}_z+{B}{\widehat{\sigma}_x}$, whose quantum counterpart is used to model the dynamics of quantum nanowires.
A similar situation appears in the second case under consideration, that is the case of \emph{pure-dephasing} Hamiltonians. These are of the type $\widehat{H}(q,p)={H}_0(q,p)+{H}_I(q,p)\widehat{A}$, where $\widehat{A}$ is a purely quantum operator. Although simple, this type of Hamiltonian is used widely in both optics and chemistry; see \cite{Manfredi} for a discussion in the context of MQC dynamics. Then, one can select a state vector that is constant in phase-space, so that $\Lambda=1$ and we obtain the mean-field equilibria
\[
(\widehat{A}-a_n)\psi_n=0
      \,,\qquad\qquad
      D_n= \frac{e^{-\mu( H_0+ H_Ia_n)}}{{\int}e^{-\mu ( H_0+ H_Ia_n)}\de q \de p}.
\]
A similar result applies whenever $\widehat{H}(q,p)$ is diagonalized by a matrix that is independent of the phase-space coordinates. 
We remark that, in the case ${\partial_pH_I=0}$, this type of mean-field equilibria coincides with those obtained by simple Born-Oppenheimer molecular dynamics.

The treatment above can be easily extended to the Uhlmann representation in such a way to recover different types of equilibria. In this case, we consider the following maximum-entropy principle:
\begin{multline*}
\delta\left[ {\operatorname{Tr}}{\int} DWW^\dagger\ln (DWW^\dagger/\Lambda)\de q\de p+\mu{\left({\operatorname{Tr}}{\int} DWW^\dagger\widehat{H}\de q\de p-E\right)}\right.
\\
\left.
+{\int}D\lambda_1(\|W\|^2-1)\de q\de p+{\lambda_2}{\left({\int} D\de q\de p-1\right)}\right]=0,
\end{multline*}
where we recall the  expression of the Liouville measure $\Lambda=1+\hbar{\operatorname{Im}\operatorname{Tr}}{\{W^\dagger{,}W\}}$. Then, taking variations with respect to $D$   yields
\[
\frac{D}\Lambda= \frac{e^{-  \langle W,(\mu
    \widehat{H}+{\ln} (WW^\dagger))W
    \rangle}}{Z_C}
    \,,\qquad\ 
    Z_C={\int}\Lambda e^{-  \langle W,(\mu
    \widehat{H}+{\ln} (WW^\dagger))W
    \rangle}\de q \de p,
\]
while the variations with respect to $W$ lead to
\[
      \Lambda{(
      \mu\widehat{H}
      {+}{\ln}(WW^\dagger){+}
        {\lambda_1}
      )}
      W=
      -i\hbar
      \{\langle W,(\mu
    \widehat{H}+{\ln} (WW^\dagger))W
    \rangle ,W 
      \}.
\]

In the  particular case when the MQC Hamiltonian depends uniquely on a phase-space function $\zeta(q,p)$, so that $\widehat{H}=\widehat{H}\circ\zeta$ and $\Lambda=1$, we obtain the equilibrium
\[
WW^\dagger=\frac{e^{ -\mu \widehat{H}}}{{\operatorname{Tr}}(e^{ -\mu \widehat{H}})}
            \,,\qquad\qquad
      D=\frac{{\operatorname{Tr}}(e^{ -\mu \widehat{H}})}{{\operatorname{Tr}}{\int}e^{ -\mu \widehat{H}}\de q \de p}.
\]
Then, the expression
\[
\widehat{\cal P}=DWW^\dagger=\frac{e^{ -\mu \widehat{H}}}{{\operatorname{Tr}}{\int}e^{ -\mu \widehat{H}}\de q \de p}
\]
of the MQC density operator coincides with those considered by other authors \cite{Alonso,Tully}. However, this expression fails to identify equilibrium states in the case of a general MQC Hamiltonian $\widehat{H}(q,p)$. Indeed, the identification of  equilibrium profiles in the mixed-state representation seems  difficult even for simple pure-dephasing Hamiltonians.

\section{Beyond Ehrenfest dynamics\label{beyond}}
While the Ehrenfest equations \eqref{EhrMod1} are the only consolidated MQC model satisfying a series of stringent consistency criteria, they  fail in capturing quantum dynamics with sufficient  accuracy. Recently, we used Koopman wavefunctions in classical mechanics to propose the following model beyond Ehrenfest dynamics:
\beq
i\hbar\partial_t\widehat{\cal P}+i\hbar\operatorname{div}(\widehat{\cal P}\boldsymbol{\cal X})=\big[\,\widehat{\!\mathscr{ H}},\widehat{\cal P}\big],
\label{HybEq1}
\eeq
with
\[
\boldsymbol{\cal X}=
\langle\bX_{\widehat{H}}\rangle+\frac1D{\operatorname{Tr}}\big( 
\bX_{{\widehat{H}}}{\cdot}\nabla \,\widehat{\!\boldsymbol\Sigma}
-
 \,\widehat{\!\boldsymbol\Sigma}{\cdot}\nabla\bX_{{\widehat{H}}}
\big), 
\qquad\qquad
 \widehat{\!\boldsymbol\Sigma}=\frac{i\hbar}{2D}\big[\widehat{\cal P},\bX_{\widehat{\cal P}}\big]
,
\] 
and
\[
\widehat{\!\mathscr{ H}}= \widehat{H}+\frac{i\hbar}D\big[\nabla\widehat{\cal P}-\widehat{\cal P}\nabla{\ln}\sqrt{D},\bX_{\widehat{H}}\big]
.
\] 
Despite its formidable appearance, a particle code based on   \eqref{HybEq1} was recently shown to capture peculiar quantum and classical features with accuracy levels  unachievable by Ehrenfest dynamics \cite{Bauer}.

Importantly, the system \eqref{HybEq1} has exactly the same Poisson bracket structure as in \eqref{bracket_candidate_rho}, although its Hamiltonian  functional  
\[
h(\widehat{\cal P})=\operatorname{Tr}\int(\widehat{\cal P}\widehat{H}{+} \,\widehat{\!\boldsymbol\Sigma}{\cdot}\bX_{{\widehat{H}}})\,\de q\de p
\] 
carries an additional term to the Ehrenfest energy. As a result, the same MQC Shannon entropy \eqref{ent2} applies equally to the model \eqref{HybEq1} and the same holds for its R\'enyi extension \eqref{MQCRenyi}. The investigation of MQC entropies in this  general context is left for future work.


\begin{credits}
\subsubsection{\ackname} We are grateful to Denys Bondar, Jes\'us Clemente Gallardo, Ignacio Franco, Raymond Kapral, Marcel Reginatto, Andrea Rocco, Jeremy Schofield, and Darryl Holm for several stimulating conversations. We are also indebted with Delyan Zhelyazov for assisting with some of the final calculations. The work of CT and FGB was supported by the Leverhulme  Grant RPG-2023-078. DMC acknowledges partial financial support of Grant PID2021-123251NB-I00 funded by MCIN/AEI/10.13039/ 501100011033 and by the European Union, and of Government of Aragon Grant E48-23R. FGB was partially supported by a startup grant from Nanyang Technological University.

\subsubsection{\discintname}
The authors have no competing interests. 
\end{credits}

%
%
%
%

\end{document}